# Stock Market Prediction using Natural Language Processing - A Survey

Om mane and Sarvanakumar kandasamy

Department of Computer Science and Engineering,
Vellore Institute of Technology, Vellore, India

## Abstract

*The stock market is a network which provides a platform for almost all major economic transactions. While investing in the stock market is a good idea, investing in individual stocks may not be, especially for the casual investor. Smart stock-picking requires in-depth research and plenty of dedication. Predicting this stock value offers enormous arbitrage profit opportunities. This attractiveness of finding a solution has prompted researchers to find a way past problems like volatility, seasonality, and dependence on time. This paper surveys recent literature in the domain of natural language processing and machine learning techniques used to predict stock market movements. The main contributions of this paper include the sophisticated categorizations of many recent articles and the illustration of the recent trends of research in stock market prediction and its related areas.*

## Keywords

*Stock market Prediction, Sentiment Analysis, Opinion Mining, Natural Language Processing, Deep Learning.*

## 1. Introduction

Stock movement prediction is a central task in computational and quantitative finance. With recent advances in deep learning and natural language processing technology, event-driven stock prediction has received increasing research attention [12, 13]. With the increasing influence of the stock market on economic trends, forecasting the trend of stocks has become a hot topic in research. Many researchers have conducted scientific and meticulous research on the stock market, trying to formulate rules for the operation of the stock market. However, the results of the research have found that the changes in the stock market seem to be unrelated [14, 15] The efficient market hypothesis theory proposed by Eugene Fame is a more authoritative explanation in the current financial circles to study the law of stock market changes. In this theory, the stock price is mainly affected by future information, namely news, rather than being driven by current or past prices [16-18]. For example, Lin et al. proposed an end-to-end hybrid neural network, which uses convolutional neural networks (CNNs) to extract data features and uses long- and short-term memory recurrent neural networks to capture the long-term dependence in the historical trend sequence of the time series to learn. Contextual features predict the trend of stock market prices [19]. Hu et al. designed a hybrid attention network (HAN) to predict stock trends based on related news sequences [20]. Li et al. proposed a multitask recurrent neural network (RNN) and a high-order Markov random domain to predict the movement direction of stock prices Hindawi Security and Communication [21]. The most popular technique used in financial time series analysis to detect the trend and seasonality is Autoregressive integrated moving average. ARIMA. The current advancement in technology has improved modern techniques in forecasting financial time series data. Deep LSTM, Shallow LSTM, 1-CNN and machine learning models were used to predict stock market data. An attention LSTM model is also used in prediction of financial time series data. ARIMA model is used in prediction of the closing prices of time series data. Va Yassine, Khadija and Faddoul in formulated classification model designed based on LSTM network to predict the probability of investing or not. For each of the selected papers the extracted data and information about (a) research characteristics, as authors, year of publication, languages covered, methodology, corpus characteristics; (b) sentiment level and categorization (binary, ternary, or fine-grained, e.g., rates $0 - 5$); (c) deep learning architectures and techniques, and (d) results and effectiveness of the proposal against baselines or state-of-the-art models.

This survey can be useful for newcomer researchers in this field as it covers the most famous Sentiment Analysis (SA) techniques and applications in one research paper. This survey uniquely gives a refined categorization to the various SA techniques which is not found in other surveys. Existing work has investigated news representation using deep learning [4, 8], neural fuzzy networks [2], heterogeneous graphs [7], long short-term memory and the random forest framework [10].

## 2. RELATED WORKS

### 2.1. Noisy Recurrent State Transition

Heyan Huang, Xiao Liu, discussed the prediction of the stock market through news events with the help of advances in deep learning techniques and machine learning technology. They proposed a novel future event prediction module to factor in likely next events according to natural events consequences. The future event module was trained over "future" data over historical events. They were able to visualize past events over a large time window, and their model is also more explainable. To their knowledge, they are the first to explicitly model both events and noise over a fundamental stock value state for news-driven stock movement prediction.

### 2.2. Deep Learning

Jiake Li, tried to establish a stock index forecast and network security model based on time series and deep learning. Based on the time series model, the author proposed to use CNN to extract in depth emotional information to replace the basic emotional features of the emotional extraction level. At the data source level, other information sources, such as basic features, are introduced to further improve the predictive performance of the model. The prediction model constructed in this paper is also based on the model in the specific direction of deep learning in data mining. Agüero, M.M. Salas, proposed the method of Multilingual Sentiment Analysis (MSA) which is an attempt to address the sentiment analysis issue through several strategies. They improved over previous reviews with wider coverage from 2017 to 2020 as well as a study focused on the underlying ideas and commonalities behind the different solutions to achieve multilingual sentiment analysis.

### 2.3. Constrained Learning

Chen, X., Rajan, D., & Quek, H.C, proposed an efficient and interpretable neuro-fuzzy system for stock price prediction using multiple technical indicators with focus on interpretability–accuracy trade-off. Neuro-fuzzy systems offer to represent complex solutions in a natural language like representation in the form of interpretable fuzzy rules which are easier to comprehend by users.

These systems are well suited to uncertain and complex real-world problems like stock market environments as these systems can deal with incomplete and uncertain data conditions and do not require large numbers of observations as other AI-based prediction models. Various studies have integrated neuro-fuzzy systems with other techniques to improve the prediction accuracy. Atsalakis et al. (2011) introduced a novel stock trading system based on neuro-fuzzy modeling method. The trading system augmented the neuro-fuzzy system with the concepts from Elliot Wave Theory to enhance forecasting accuracy. The study demonstrated high performance of this system in predicting the trends in stock prices. Tan et al. (2011) presented an ANFIS-based model for stock trading. Authors supplemented the model using reinforcement learning (RL).

## 2.4. Graph based Prediction

Junran Wua, Ke Xua discussed a method to develop a novel framework based on the structural information embedded in price graphs by conducting numerous experiments on real-world market data. In this study, they propose a novel framework to address both issues. Specifically, in terms of transforming time series into complex networks, they convert market price series into graphs. Then, structural information, referring to associations among temporal points and the node weights, is extracted from the mapped graphs to resolve the problems regarding long-term dependencies and the chaotic property. They take graph embeddings to represent the associations among temporal points as the prediction model inputs. Node weights are used as a priori knowledge to enhance the learning of temporal attention. The effectiveness of their proposed framework is validated using real-world stock data, and their approach obtains the best performance among several state-of-the-art benchmarks. Feature based methods only use simple lexical features such as bags-of-words, noun phrases and named entities (Kogan et al., 2009; Schumaker and Chen, 2009) extracted from financial text to predict the movement of stock market. To better predict the movement of stock market, Ding et al. (2015, 2019) propose to utilize the event level information extracted from titles or abstracts. Moreover, to capture contextual information within document, many document representation learning based methods (Augenstein et al., 2016; Chang et al., 2016; Tang et al., 2015; Yang et al., 2016) are proposed to learn a distributed representation of the whole document or abstract, and build prediction model based on the representation vector, so that information from overall document can be utilized for making predictions.

## 2.5. Liquidity Prediction for Learning Models

As an evaluative consideration in investment decisions, stock liquidity is of critical importance for all stakeholders in the market. It also has implications for the stock market's growth. Liquidity sanctions investors and issuers to meet their quotas regarding investment, financing or hedging, reducing investment costs and the cost of capital experimental results, it can be concluded that the LSTM model allows for prediction characterized by lowest value of mean square error (MSE). The aim of [6] is to develop the machine learning models for liquidity prediction. The subject of research is the Vietnamese stock market, focusing on the recent years - from 2011 to 2019. Vietnamese stock market differs from developed markets and emerging markets. It is characterized by a limited number of transactions, which are also relatively small. The Multilayer Perceptron, Long-Short Term Memory and Linear Regression models have been developed. Data was prepared for further analysis before data was preliminarily analyzed and relationships were found that enable illiquidity prediction. After preparing and selecting data, a deep learning model was developed. The study was conducted on the Ho Chi Minh City Stock Exchange in Vietnam. A sample of daily data of 220 companies representing different sectors was used for analysis. The main disadvantage of the proposed approach is the performing prediction only for one day.

## 3. ARCHITECTURE

This section gives a detailed explanation about the methods used in analyzing and evaluating the trend of S&P500 stocks, where a System Architecture of stock market price prediction is designed which explains its associated process flow. Also, there has been a great interest in novelty architectures for Sentiment Analysis (SA). One approach that has received substantial attention when working at the basic document level, is the design of hierarchical models which learn a representation for sentences from its words, on top of this level, another model can learn representations for documents. Other substitutes such as Convolutional Neural Networks (CNN) or Long Short-Term Memory (LSTM) can be used at each level. Works in [25,26] and [27] are some examples of this approach.

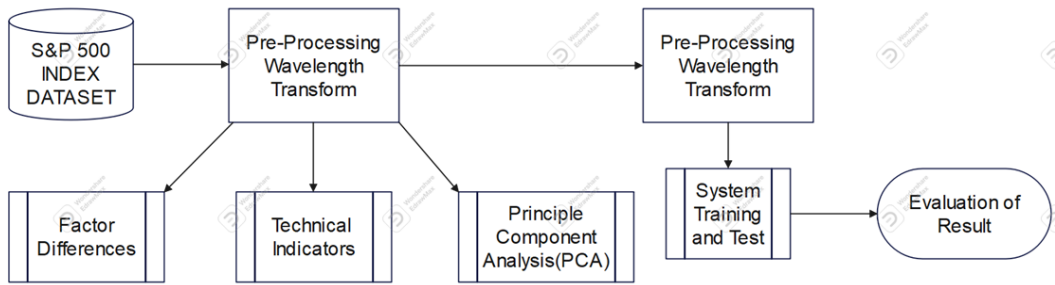

Figure 1. System Architecture [3]

In [1] the authors explicitly model both noise and events over a recurrent stock value state, which is modelled using LSTM. For each day of trading, they consider the news events happened in that day as well as the past news events using neural attention [28]. Assessing the impacts of insider trading, they also pre-suppose future news in the training method and procedure. To model the high stochasticity of stock market, they illustrate a supplemented noise using a neural module. Their model is named attention-based noisy recurrent states transition (ANRES) as can be seen from figure 2.

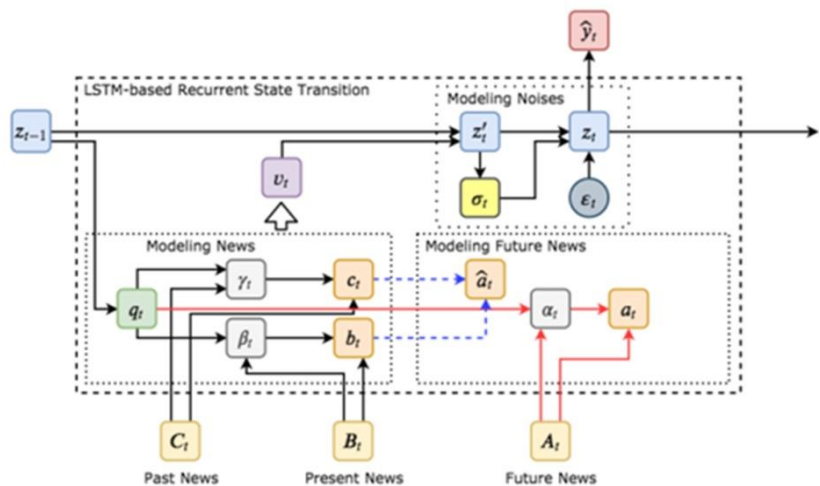

Figure 2. The ANRES model framework for trading day t in a trading sequence. The black solid elbows are used both in the training and the evaluating procedures. The red solid elbows are only used in the training procedure, while the blue dotted elbows in the evaluating procedure [1]

## 3.1. Time Series Model

The object of the stock model based on time series is the historical data of stocks. The core step is to divide the historical data of stocks to facilitate the subsequent stock market forecasts. In this model, the first and most important step is to collect and process time series data. When predicting a time series, it is mainly by observing the trend changes of the time series first and predicting future time series changes by learning the law of past changes. Time series data often have a large amount of data and are difficult to process directly. This requires dividing it and dividing the time series by finding the key trend points. Through this division method, the originally complex data can be compressed while also removing some noise in the stock sequence. Some points that are not helpful for prediction, so that the retained information is more effective for the model to learn the changes in the time series data, and the time series rules can be found more clearly [4].

## 3.2. Deep Learning Model

First, for traditional classifiers (such as SVM and KNN) to deal with the general problem of time series data classification, with the help of the recurrent neural network to facilitate the modelling of time series data, a depth-based stock prediction model learned, and on the basis of this model, the sentiment analysis results of stock-related data in the social media text are added to construct a trend prediction model that integrates basic emotional features. Among the deep learning technologies that have emerged in recent years, convolutional neural networks are the most widely used. Figure 3 shows the index prediction process based on deep learning [4].

## 3.3. Heterogeneous Graphs

Given one or several news documents of a corporation, [7]'s goal is to predict the future stock reaction of this corporation based on corresponding financial text. Rather than predicting the specific price reaction range, we formulate the prediction as fluctuation polarity. The prediction will be "Positive" for the rise in stock price or "Negative" for the decline in stock price. The edges in the heterogeneous graphs are undirected.

Table 1. Example of sentences in Financial Text and stock prices of corporations

| | |
|---|---|
| *Sentence* | *After three consecutive days rise on stock prices, H & M was pushed to the forefront of public opinion due to the Xinjiang cotton incident.* |
| *Stock Price Reaction* | *H&M Negative* |

## 4. METHODOLOGY

### 4.1. Dataset Description

To study the proposed algorithm, Hadi Rezaei, Hamidreza Faaljou extract the historical daily financial time series from January 2010 to September 2019 from the Yahoo Finance Website. The daily data includes the close price of S&P 500, Dow Jones, DAX, and Nikkei225. The authors use the dataset released by Duan et al. in their experiments. The dataset includes more than 100k financial articles from Reuters. They have used an open source openIE tool for information extraction. Training is stopped if the performance doesn't improve for 10 epochs.

Use of heuristic rules to build connections among words, event triples and sentences is done. Employment of multi-grained encoders to encode the nodes within HGk into embedding is done. Following baselines are used in this model: Sentiment, Event Tensor, Event Tensor-CS, TGT-CTX-LSTM, Conditional Encoding, TGT-HN, TGT-HAN, GCN, GAT. The paper [6] covers the sample period from January 2011 to December 2019, which contains 2,242 trading days. The eligible stocks in the sample consist of 378 companies in two stock exchanges, which included 179 stocks on the HOSE and 199 stocks on the HNX. Stock market liquidity is represented by seven liquidity measures.

[6] Covers the sample period from January 2011 to December 2019, which contains 2,242 trading days. The eligible stocks in the sample consist of 378 companies in two stock exchanges, which included 179 stocks on the HOSE and 199 stocks on the HNX. In [2] Daily stock data of three indices, viz. BSE, CNX Nifty and S&P 500, comprising of five fundamental stock quantities, namely maximum price, open price, minimum price, close price, and trading volume, are used to evaluate the proposed system. The dataset of BSE index is from January 30, 2005, to December 30, 2015. For Nifty index, dataset from January3, 2005, to December 24, 2015, is used. In case of S&P 500, daily stock data of 2865 records from February 9, 2005, to June 28, 2016, have been used.

## 4.2. Dataset Pre-Processing

To work with this dataset, there is a need to pre-process it, wherein visualization followed by illustration is foremost importance. This results in us being able to identify trends, missing values, noise, and outliers. To study the proposed algorithm, data was extracted from the historical daily financial time series from October 2021 to April 2022 from the Yahoo Finance Website. The daily data includes the close price of S&P500, Dow Jones and DAX. To better visualize these indices, The data is described in graph 1.

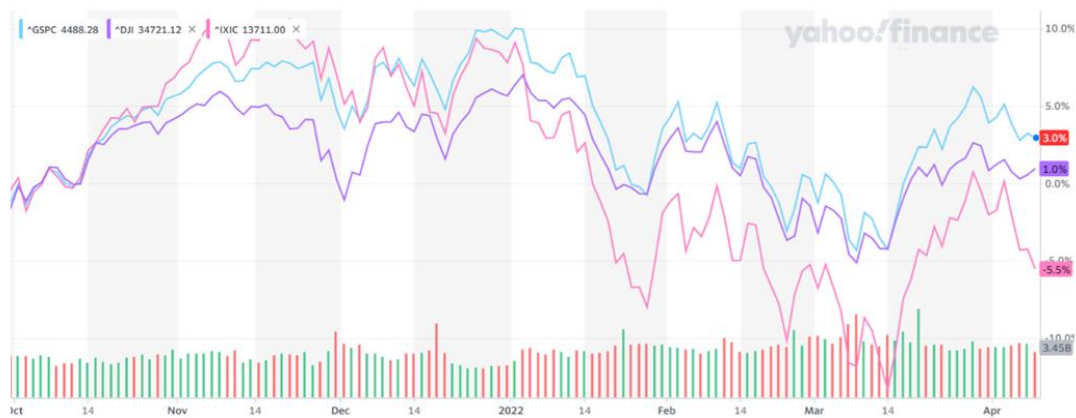

Graph 1. Price Index for S&P 500, Dow Jones, DAX

Due to the complexity of financial stock exchange environment, stock prices contain a lot of noise that make it very difficult to achieve a good model accuracy when trying to forecast it trend movement. Since is the interest of investors to achieve good forecasting model, then there is a need to reduce the noise in the dataset because stock environment is containing noise from the news articles and other source media information. A wavelength transform, a mathematical function introduced in the pre-processing stage by de noise S&P500 dataset and present it trend and structure of the dataset. The authors transform the dataset using wavelength mathematical transform function:

$$X_\omega(a,b) = \frac{1}{\sqrt{a}} \int_{-\infty}^{\infty} x(t)\psi\left(\frac{t-b}{a}\right) dt$$

After transformation we drop the co-efficient with more standard deviation away from the co-efficient and inversely transform the new co-efficient to get the denoise S&P 500 dataset [3].

## 4.3. Equations

From [1] the ANRES model uses LSTM with peephole connections. The underlying stock value trends are represented as a recurrent state z transited over time, which can reflect the fundamentals of a stock. In each trading day, they consider the impact of corresponding news events and a random noise.

$$z'_t = \overrightarrow{\text{LSTM}}(v_t, z_t - 1)$$

$$z_t = f(z'_t)$$

where $v_t$ is the news events impact vector on the trading day t and f is a function in which random noise will be integrated. In deep learning models, loss error is usually used to evaluate the prediction, which refers to the difference between the actual observed value and the predicted value. In order to evaluate the loss error, different tools can be used. The standard metrics in this type of models include root-mean square error (RMSE), mean absolute error (MAE), and mean absolute percentage error (MAPE) [8]. The formula for their calculation is as follows:

$$MSE = \frac{1}{n}\sum_{i}^{n}(y_i - \hat{y}_i)^2$$

$$MAE = \frac{1}{n}\sum_{i=1}^{n}|y_i - \hat{y}_i|$$

$$MAPE = \frac{1}{n}\sum_{i=1}^{n}\left|\frac{y_i - \hat{y}_i}{y_i}\right|$$

In order to select key information and get the stock price fluctuation of a specific corporation. Firstly, we use the corporation to softly select relevant information from sentences:

$$\alpha_i = H_c H'_s,$$

$$A_i = \frac{\alpha_i}{\Sigma_{j\epsilon S_k}\alpha_j}$$

$$u = AH_s$$

where $H_c \epsilon \mathbb{R}^d$ is the corporation representation and $H'_s \epsilon \mathbb{R}^{d*r}$ is the embedding of sentences, $A \epsilon \mathbb{R}^r$ is the weight matrix and $U \epsilon \mathbb{R}^d$ is the weight sum of sentences representation.

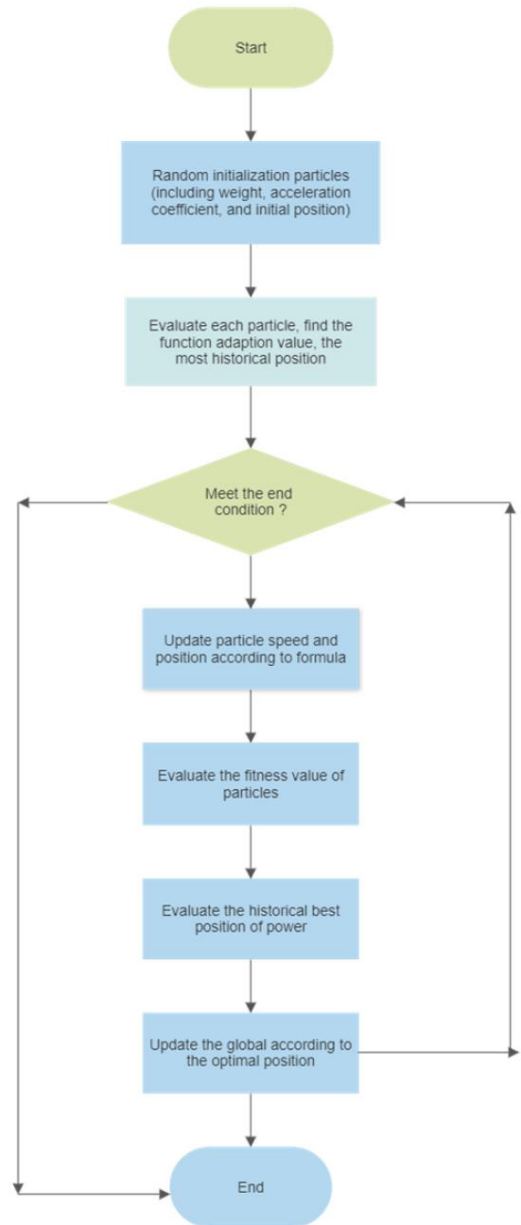

Figure 3. Index prediction process based on deep learning [4]

Therefore, based on U, the authors have used a linear function to predict future stock market fluctuation of this corporation:

$$X_{pred} = Concate[H_c, U],$$

$$Probabilities = Softmax(W_{pred} X_{pred}),$$

$$Prediction = ArgMax(Probabilities),$$

where $X_{pred} \epsilon \mathbb{R}^{2*d}$ is the concatenation of corporation and sentence representation and $W_{pred} \epsilon \mathbb{R}^d$ is a trainable parameter. Probabilities $\epsilon \mathbb{R}^2$ denotes the final distribution of Positive and Negative. Predictions are the final prediction on stock market fluctuation, 0 and 1 denotes Negative and Positive, respectively.

## 5. EXPERIMENTS AND EVALUATION

### 5.1. Evaluation Metrics

In [2] in order to evaluate the performance of the proposed model, root- mean-squared error (RMSE) and mean average percentage (MAPE) have been used to measure forecasting accuracy. RMSE and MAPE measure the deviation between the actual and the forecasted values. Smaller the value of these metrics larger is the forecasting performance of the model. Also, directional accuracy (DA) has been used which gives the correctness of a system for the predicted trend of a stock price index. For [3] MACD Moving average convergence divergence displays the trend following the characteristic of market prices. CCI Consumer channel index identifies price reversals, price extreme and the strength of rise prices. ROC Rate of change is a momentum indicator that measures the percentage change in prices from one period to the next. For comparison, the baseline approach is chosen based on classical machine learning algorithms. The machine learning algorithm is a traditional method that is widely used but less complex than the model used in this work. Using the same input variables trained and tested their system architecture predictions on S&P 500 new process data. They chose the Logistic regression and Random Forest model. They implemented the two baseline models using Scikit -Learn library. List of supporting algorithms used in this paper are RSI, Momentum, OBV, MA, EMA, ADX. In [7] the authors use the dataset released by Duan et al. (2018) in their experiments. The dataset includes more than 100k financial articles from Reuters2, which also includes stock prices of different corporations. The time interval of the articles starts from October 2006 to December 2015. Only articles that mentioned at least one firm are selected by them, and the dataset is balanced on positive and negative classes. The statistics of this dataset are listed in Table 2. Evaluation Metrics Follow the metrics used in prior works (Chang et al., 2016; Duan et al., 2018), they adopt the area under the precision-recall curve (AUC) as our evaluation metrics. To construct the heterogeneous graph, they use an open source openIE tool (Schmitz et al., 2012) for information extraction. Following the settings of Wang et al. (2020), they initialize the word nodes with 300-dimension GloVe embeddings (Pennington et al., 2014), and the dimension of every node feature is d = 128. The heads of multi-head attention is set as 8. For the training process, the learning rate is set as 5e-4. Moreover, they apply an early stopping mechanism, training will be stopped if the performance on the development set doesn't improve for 10 epochs.

Table 2. The statistics of [7] dataset.

|  | **Train** | **Development** | **Test** |
|---|---|---|---|
| Positive | 9,376 | 477 | 973 |
| Negative | 9,394 | 486 | 981 |
| Total | 18,770 | 963 | 1,954 |
| Docs | 27,657 | 1,419 | 2,989 |

For trend forecasting in financial time series with indicator system, the evaluation metric formula is as follows:

$$y_{prob} = \begin{cases} 0, \hat{y}_{model} < 0.5 \\ 1, \quad otherwise \end{cases}$$

They further move on to evaluate the performance of our prediction model, the selected performance metrics: Recall, Precision and F1 is chosen. They define the chosen metrics as positive class (true positive), negative class (true negative).

$$Precision = \frac{truepositive}{truepositive + falsepositive}$$

$$Recall = \frac{truepositive}{truepositive + falsenegative}$$

For ANRES Model in [1] the authors use the public financial news dataset released by [13], which is crawled from Reuters and Bloomberg over the period from October 2006 to November 2013. They conduct their experiments on predicting the Standard & Poor's 500 stock (S&P 500) index and its selected individual stocks, obtaining indices and prices from Yahoo Finance. Detailed statistics of the training, development and test sets are shown in Table 3.

Table 3. Statistics of ANRES Model Dataset

|  | **Training** | **Development** | **Test** |
| --- | --- | --- | --- |
| #documents | 358,122 | 96,299 | 99,030 |
| #samples | 1,425 | 169 | 191 |
| time span | 10/20/2006-06/18/2012 | 06/19/2012-02/21/2013 | 02/22/2013-11/21/2013 |

Following previous work [20, 3, 21], they adopt the standard measure of accuracy and Matthews Correlation Coefficient (MCC) to evaluate S&P 500 index prediction and selected individual stock prediction. MCC is applied because it avoids bias due to data skew. Given the confusion matrix which contains true positive(tp), false positive(fp), true negative(tn) and false negative(fn) values, MCC is calculated as:

$$MCC = \frac{tp \times tn - fp \times fn}{\sqrt{(tp+fp)(tp+fn)(tn+fp)(tn+fn)}}$$

In summary, the following four baselines are designed:

• ANRES_Sing_R: randomly initializing the states for each single trading day.
• ANRES_Sing_Z: initializing the states as zeros for each single trading day.
•ANRES_Seq_R: randomly initializing the first states for each trading sequence only.
• ANRES_Seq_Z: initializing the first states as zeros for each trading sequence only.

For Deep Learning Model, the evaluation is done by Simulation Environment and Data. Compared with individual stocks, the volatility of stock indexes is generally smaller because stock indexes are composed of many stocks in different industries and can better reflect the overall economic momentum and overall condition. Index Forecasting Effect Analysis. Using the 1219-day data samples of the Shanghai Composite Index for 5 years from 2015 to 2019, the stock data of 10 consecutive days and 20 days were used as input samples to establish a prediction model for closing price prediction. Using the 731-day data sample of the Shanghai Composite Index for 3 years from 2017 to 2019, 5 consecutive days and 10 days of stock data were used as input samples to establish a prediction model for closing price prediction.

## 5.2. Baselines

- **Sentiment** (Mayew and Venkatachalam, 2012) is a lexicon-based sentiment analysis method. We use the sentiment lexicons and method released by Loughran and McDonald (2011).
- **Event Tensor** (Ding et al., 2015) is a neural tensor network based event representation learning method. They extract event triples from titles or abstracts. The event triples from the abstract will be averaged for prediction
- **Event Tensor-CS** (Ding et al., 2019) is an external common sense knowledge enhanced event representation learning method. Similar events are trained to be close to each other in vector space by predicting their sentiment polarities and contextual events.
- **TGT-CTX-LSTM** (Chang et al., 2016) uses dependency parse tree to learn a target-related representation of abstract for stock market prediction.
- **LSTM:** a basic LSTM network used to predict the future trends of stock prices based on historical price data [29].
- **DARNN:** a dual-stage attention-based RNN that employs input attention and temporal attention in the encoder and decoder stages, respectively [30].
- **DARNN-SA:** an extension of the DARNN that employs a self-attention layer between the output of the DARNN and the prediction layer [9]
- **MFNN:** a multifilter deep learning model that integrates convolutional and recurrent neurons for feature extraction with respect to financial time series and stock prediction [31].
- **CA-SFCN:** a fully convolutional network (FCN) incorporating cross attention (CA), in which CA is also used as dual-stage attention for the variable and temporal dimensions (with temporal attention first) [32].

## 6. RESULTS

Given below are the results which were obtained after reviewing several research papers:

Table 4. Result of Deep Learning (LSTM) and Baseline models by indicator system

| Metrics | Deep Learning (LSTM model) (%) | Baseline (random forest model) | Baseline (logistic regression model) |
| --- | --- | --- | --- |
| Accuracy | 0.687% | 0.56% | 0.551% |
| Precision | 0.538 | 0.514 | 0.594 |
| Recall | 0.697 | 0.412 | 0.236 |

In heterogeneous graphs the overall results are shown in Table 5, from which we can make the following observations:

Table 5. Results of heterogenous graph method [7]

| Methods | AUC |
| --- | --- |
| Sentiment | 0.533 |
| Event Tensor + Title | 0.544 |
| Event Tensor + Abstract | 0.549 |
| Event Tensor-CS + Title | 0.564 |
| Event Tensor-CS + Abstract | 0.570 |
| Conditional Encoding | 0.603 |
| TGT-CTX-LSTM | 0.632 |
| TGT-HN | 0.615 |
| TGT-HAN | 0.633 |
| GCN | 0.621 |
| GAT | 0.615 |
| **HGM-GIF** | **0.638** |

1. Semantic knowledge acquisition methods (Event Tensor, Conditional Encoding, TGT-CTX-LSTM, TGT-HN, TGTHAN, HGM-GIF) achieve better results than feature-based method. The reason is that learning semantics within financial text is more effective than learning linguistic features for stock market prediction.

2. Comparison between Event Tensor –CS, Conditional Encoding, TGT-CTX-LSTM, TGT-HN, TGT-HAN, HGM-GIF and Event Tensor shows that external knowledge or sentences within documents can supplement contextual information and offer more useful information for prediction. This confirms that predictions need rich contextual information. With the sequential heterogeneous graph layer, we can capture rich contextual information within financial text.

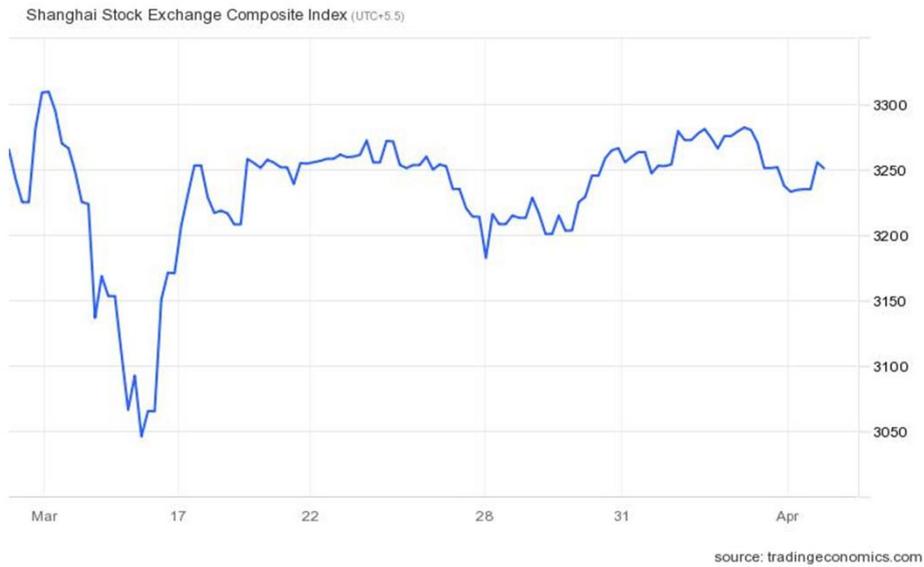

Figure 4. The prediction results of the Shanghai Composite Index at 30-days.

Using the 1219-day data samples of the Shanghai Composite Index for 5 years from 2018 to 2022, the stock data of 7 consecutive days and 30 days were used as input samples to establish a prediction model for closing price prediction. +ese two models are called SHYSD10 and SHYSD20, respectively. Figures 4 and 5 show their prediction results. Figure 4 shows the prediction results of the Shanghai Composite Index at 30-day intervals. Figure 5 shows the forecast results of the Shanghai Composite Index at 70 consecutive days.

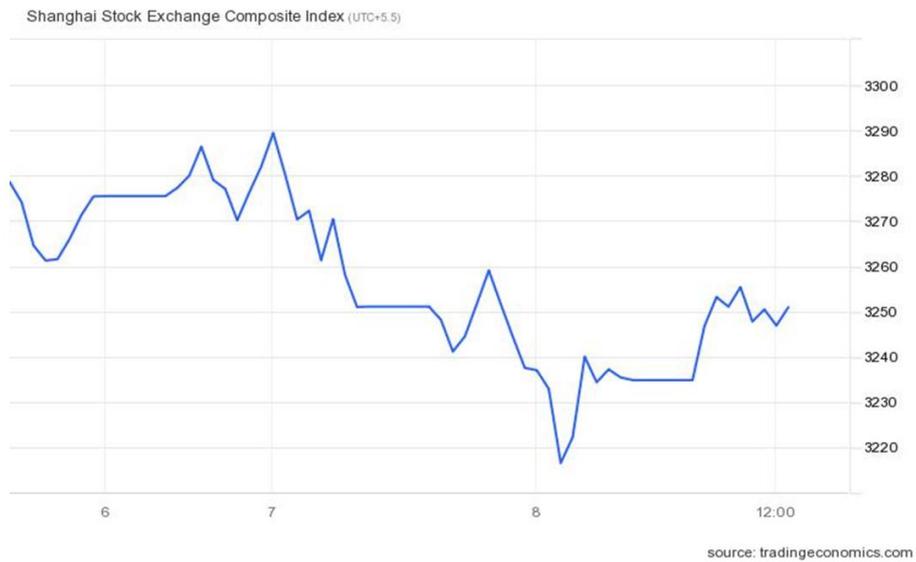

Figure 5. The prediction results of the Shanghai Composite Index at 7 consecutive days.

|  |  | Number of Hidden Units | | RMSE | MAE | MAPE |
|---|---|---|---|---|---|---|
|  |  | LSTM | CNN |  |  |  |
| S&P500 | EMD-CNN-LSTM | 200 | 512 | 14.88 | 12.04 | 0.611 |
|  | EMD-LSTM | 200 | -- | 15.51 | 12.60 | 0.639 |
| Dow Jones | EMD-CNN-LSTM | 200 | 512 | 163.56 | 120.97 | 0.6729 |
|  | EMD-LSTM | 200 | -- | 171.40 | 128.55 | 0.7184 |
| DAX | EMD-CNN-LSTM | 200 | 512 | 108.56 | 86.05 | 0.907 |
|  | EMD-LSTM | 200 | -- | 109.97 | 86.75 | 0.920 |
| Nikkei225 | EMD-CNN-LSTM | 200 | 512 | 194.17 | 147.18 | 0.9413 |
|  | EMD-LSTM | 200 | -- | 213.45 | 164.08 | 1.0513 |

Figure 6. Prediction based on EMD based algorithm

## 7. CONCLUSION

Concerning the significance of stock market prediction and its challenges, researchers always try to introduce modern methods in the analysis of these markets. LSTM as a state-of-the-art model and CNN which are deep learning models yield good results in the analysis of stock market. EMD and CEEMD are among the effective algorithms that have recently been considered in this research area. Thus, this article sought to introduce the proposed algorithm of CEEMD-CNN-LSTM by studying each of these models and evaluate them based on data of different stock price indices. The major concept of the suggested algorithm was to create a collaboration between CEEMD, CNN, and LSTM models by joining them together, which could extract deep features and time sequences. Then, we applied the trained algorithm for one-step-ahead stock price forecasting. In this regard, first, we decomposed the financial time series to different IMFs and the residual by CEEMD and EMD algorithms. The IMFs and the residual were separately analysed by CNN-LSTM model. Facilitation of their analysis with CEEMD and EMD, as well as extracting features and patterns in data with CNN and further analysis in the context of the time plus analysis of the dependencies with LSTM enhanced the predictive capabilities of this model. The practical results of this article also support this claim. Note that the assessment metrics used in this study were RMSE, MAE, and MAPE [3].

## 8. FUTURE WORKS

ANRES model tends to pay more attention to a new event when it first occurs, which offers us a potential improving direction in the future. In future, the model can be applied to other stock indexes like NSE, TAIEX. For further improving the system accuracy, an underlying TSK-based fuzzy system with a constrained gradient-based technique can be employed. Also, a constrained gradient-based technique can be integrated with a Mamdani-type fuzzy system so that interpretability is not compromised which may happen in case of a TSK system. For reducing the rule base size, techniques like similar rule merging can also be used. Some area that will be needed to achieved better accuracy in our future work, some scholars are concerning the used of sentiment analysis on twitter to predict stock market trend movement, Beside the social media, other qualitative indicators like news, internet and domestic policy changes can also be used as input to predict the trend of stocks. Another important concept is Elliot waves principles which can perform better on stock market trend prediction.

**ACKNOWLEDGEMENTS**

The authors would like to thank Vellore Institute of Technology for providing them with the resources necessary to complete this paper.

## AUTHOR

**Om Mane** is a sophomore at Vellore Institute of Technology, Vellore in Computer Science and Engineering.

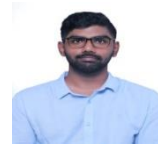